\documentclass[aps,pre,reprint,superscriptaddress,amsmath,showpacs,amssymb]{revtex4-1}
\usepackage{graphicx}
\usepackage{tikz}
\usetikzlibrary{backgrounds}
\usepackage{amssymb,latexsym}
\usepackage{amsfonts}
\usepackage{color}
\usepackage{epsfig}
\usepackage{graphicx}
\usepackage{pgfplots}
\usepackage{empheq}

\bibstyle{apsrev.bib}

\newcommand{\caA}{{\mathcal A}}
\newcommand{\caB}{{\mathcal B}}

\newcommand{\caF}{{\mathcal F}}

\newcommand{\caT}{{\mathcal T}}

\newcommand{\diff}{\text{d}}

\newcommand{\mean}[1]{{\left< #1 \right>}}
\begin{document}

\title{Interacting Brownian dynamics in a nonequilibrium particle bath}

\newcommand\uleip{\affiliation{Institut f\"ur Theoretische Physik, Universit\"at Leipzig,  Postfach 100 920, D-04009 Leipzig, Germany}}
\newcommand\MPI{\affiliation{Max Planck Institute for Mathematics in the Sciences, Inselstr. 22, 04103 Leipzig, Germany}}

\author{Stefano Steffenoni}
\email{stefano.steffenoni@mis.mpg.de}\MPI\uleip
\author{Klaus Kroy}
\email{klaus.kroy@itp.uni-leipzig.de}\uleip
\author{Gianmaria Falasco}
\email{falasco@itp.uni-leipzig.de}\uleip


\date{\today}

\begin{abstract}
% The generalized Langevin equation for many probe particles weakly
% interacting with a driv%en environment \textbf{(fluid?solvent?)} is
% derived by applying nonequilibrium linear respo%nse theory.

We set up a mesoscopic theory for interacting Brownian particles
embedded in a nonequilibrium environment, starting from the
microscopic interacting many-body theory.  Using nonequilibrium linear
response theory, we characterize the effective dynamical interactions
on the mesoscopic scale and the statistics of the nonequilibrium
environmental noise, arising upon integrating out the fast degrees of
freedom. As hallmarks of nonequilibrium, the breakdown of the
fluctuation-dissipation and action-reaction relations for Brownian
degrees of freedom are exemplified with two prototypical models for
the environment, namely, active Brownian particles and stirred
colloids.

\end{abstract}

%\begin{abstract}
%The generalized Langevin equation for many probe particles weakly interacting with a driven environment \textbf{(fluid?solvent?)} is derived by applying nonequilibrium linear response theory. 
%We characterized the effective dynamical interactions between the probes, as well as the noise statistics, which arise integrating out the fast degrees of freedom of the environment. The hallmarks of nonequilibrium, i.e. the breakdown of the fluctuation-dissipation theorem and the action-reaction law for the effective forces between the probes, are exemplified considering prototypical nonequilibrium environments, namely, active Brownian particles and stirred colloids.
%
%\end{abstract}

\maketitle

\section{Introduction}
The notion of Brownian motion refers to the thermal fluctuations of
some mesoscopic particles in contact with a bath of smaller
particles. Colloidal beads dissolved in a simple fluid are the
historical paradigm. But the concept generalizes to any slow
mesoscopic degrees of freedom in contact with a bath of fast
microscopic degrees of freedom. In fact, there are wide-ranging
applications of the basic theme outside the realm of physics~\cite{fre05}. The essential feature is a scale separation
between the Brownian and bath degrees of freedom that allows for some
systematic coarse-graining of an otherwise intractable many-body
system. A convenient approach to formalize this seminal insight is via
(generalized) Langevin equations, which can be formulated for a wide
variety of phenomena and have helped to rationalize a range of
interesting phenomena from long-time tails
\cite{mclennan.1989} to critical fluctuations
\cite{hoh77}. They have therefore become a prevalent tool in the
quantitative description of soft matter and, more generally, noisy
systems. 

Even though there exist systematic derivations of such
mesoscopic equations of motion from an underlying microscopic
many-body Hamiltonian through the elimination of the fast degrees of
freedom \cite{zwa73, kaw73, van86}, one eventually
typically appeals to equilibrium statistical mechanics in order to
make the formal expressions practically useful.  Namely, to bypass the
explicit solution of the microscopic dynamic equations, the ``noise''
fluctuations that agitate the mesoscopic degrees of freedom are
assigned a weight in accord with Boltzmann's principle
\cite{ons53}. By construction, their correlations then satisfy
detailed balance in the form of a fluctuation-dissipation theorem
\cite{kub66}. This implies, in particular, that they induce mesoscopic
correlations in accord with equipartition. Moreover, the average
mesoscopic dynamics is found to be a gradient flow in a convex free
energy landscape.  Being derived by such a (thermodynamic) potential,
the mean effective interactions of the Brownian degrees of freedom
themselves obey the action-reaction principle. In other words, in
equilibrium stochastic thermodynamics, the symmetries holding on the
microscopic level can essentially be lifted up to the mesoscopic
scale. The theory remains valid even when some of these mesoscopic
degrees of freedom are externally driven out of equilibrium, as long
as local detailed balance persists \cite{kat83, har06}, i.e., under the
assumption that the source of non-equilibrium does not appreciably
affect the (many) bath degrees of freedom. For this reason, the
concept of a Brownian scale separation, as embodied in the Langevin
equation, has played a central role in the development of a framework
of stochastic thermodynamics that reaches out to conditions far from
equilibrium \cite{sek10, sei05} and in the study of nonequilibrium
fluctuation and work relations \cite{sei12}.

In contrast, none of the above symmetry properties generally survives
on the Brownian scale if the bath itself is driven out of equilibrium.
Not only is the detailed balance of the Brownian degrees of freedom
then lost, but also equipartition gives way to a more complex energy
partition rule \cite{fal16}, stochastic forces are no longer of
gradient-type \cite{bas15}, and the action-reaction principle is
violated \cite{ivl15}.  In soft matter physics, one finds many
examples for interacting probes in non-equilibrium baths. One may
naturally think of a suspension of colloids immersed in a
non-equilibrium solvent, such as a sheared fluid \cite{dzu03}, a
granular \cite{can09}, glassy \cite{abo04} or active-particle
suspension \cite{che07}, or even the cytoplasm of a living cell
\cite{lau03}. It would certainly be of great interest to establish a
self-contained coarse-grained description for the colloids in such
situations.  Yet, the usual equilibrium arguments invoked in the
construction of a coarse-grained Langevin description, are not any
more applicable. So the reduced stochastic description (assuming it
still exists) must be found by other means, in the worst case by
\textit{explicitly} integrating out the dynamics of the nonequilibrium
environment. It should go without saying that, for scientifically or
technologically interesting systems, this is almost always an
impossible task.

There is thus great interest in defining suitable conditions and
finding general approximate methods \cite{spo12, van85} that allow for reliable and useful
predictions on the Brownian scale, even if the microscopic degrees of
freedom of the environment are driven far from equilibrium. Among
other things, such methods should enable us to infer the key
properties of the nonequilibrium environment from the observed
mesoscopic dynamics. Ideally, they should moreover help to unravel the
formal structure of the coarse graining, such that we can identify the
mechanisms underlying the emergent violations of the detailed-balance
and action-reaction principles, which are not always straightforwardly
discernible on the mesoscopic scale. Finally, this would allow the
logical chain of arguments used to infer environmental conditions from
Brownian dynamics to be reversed; namely, to tailor some desired
mesoscopic properties by a fine-tuning of the nonequilibrium driving
of the micro-statistics of the environment, with some obvious
technological implications.

A good candidate for such an approximate method for bypassing the
integration of the microscopic dynamics is suggested by the theory of
Brownian motion, itself. If the ``fast'' bath degrees of freedom of
some Brownian system themselves admit a coarse-grained description by
a mesoscopically driven (generalized) Langevin theory routed in its own
equilibrium bath, the resulting theory fulfills all of the above
requirements. An example for a successful implementation of such a
scheme is provided by the theory of nonequilibrium fluctuating
hydrodynamics \cite{Sengers.2006}, on which theories of Brownian dynamics in
nonequilibrium baths can be based \cite{fal14,fal16b}.

\begin{figure}[t]
\center
\includegraphics[width=0.45\textwidth]{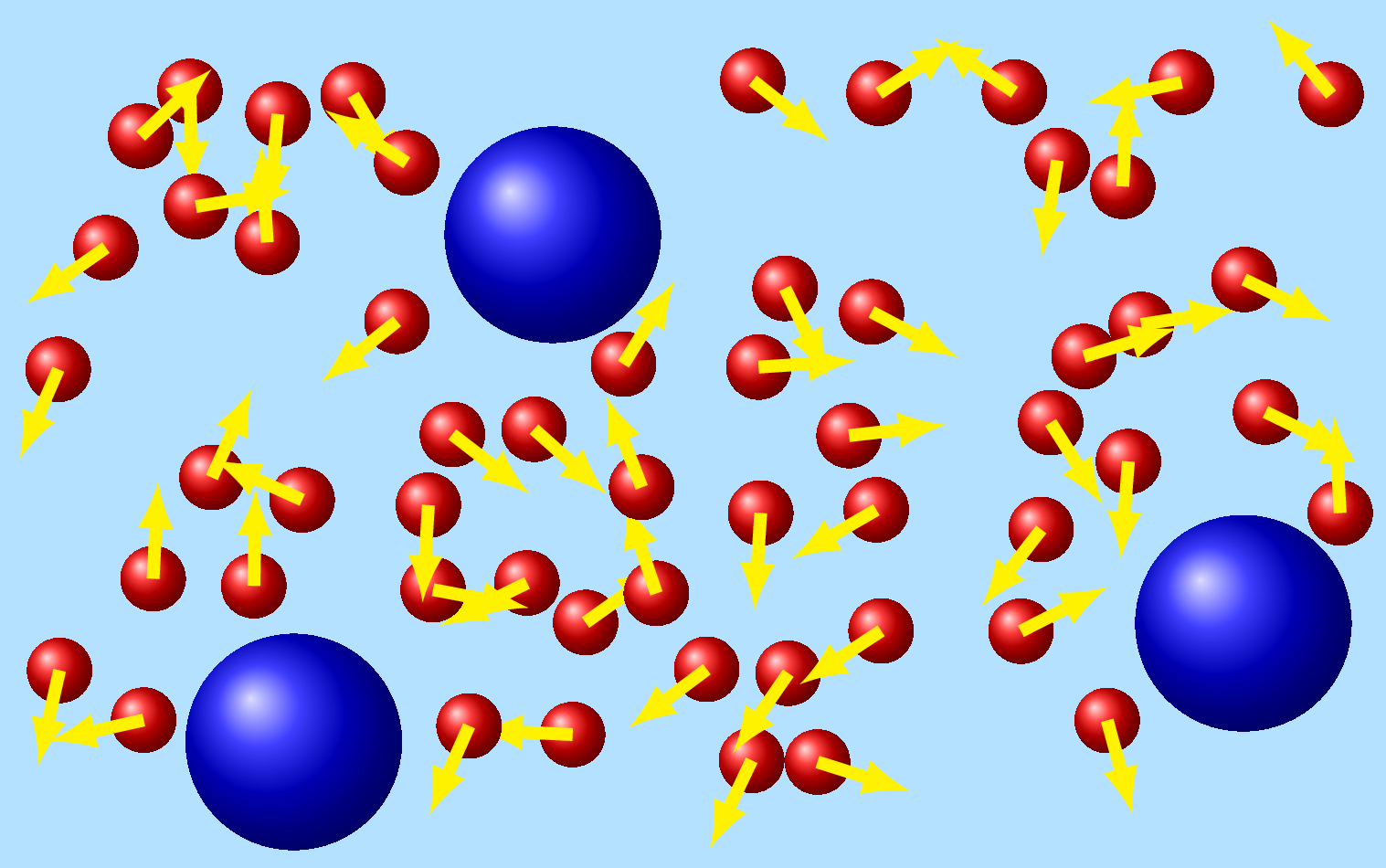}
\caption{Schematic representation of the three-level scheme employed by our theory: the probes (blue) representing the system are embedded in a nonequilibrium environment, e.g. a fluid of smaller driven particles (red), which are in contact with a stochastic equilibrium thermal bath (light blue).}
\label{fig:scheme}
\end{figure}
In the following, we pursue this idea in a slightly more abstract and
fully particle-based framework, i.e., without appealing to a
hydrodynamic limit for the environment degrees of freedom. We do
assume, instead, that the environment is made up of some sort of
particles that evolve according to some driven Markovian stochastic
dynamics enjoying local detail balance. In contrast to standard
Brownian dynamics, we thus do not require a direct buffering of the
probe degrees of freedom by some equilibrium thermal reservoir but
only an indirect one, mediated by the non-equilibrium environment
(cf.\ Fig.~\ref{fig:scheme}).  Technically, we employ non-equilibrium
linear response theory
\cite{nak08,che08,spe09,sei10,pro09,ver11a,lip05,bai09,bai09b},
to derive a Langevin equation for the interacting probe particles that
we assume to be weakly coupled to the interacting many-body system
acting as the environment.  Following \cite{mae14b, mae15}, we then go beyond
a merely static description that would only account for systematic
probe interactions induced by the nonequilibrium environment, such as
nonequilibrium depletion forces \cite{sri12, har14}. We explicitly look
for the fluctuations of such induced forces around their average
values.  In Sec.~\ref{sec:theory} we obtain formal expressions for
these fluctuating forces, the friction, and the noise statistics.
When the driving is off, we retrieve the expected detailed balance
condition connecting the noise correlation to the friction memory
kernel. But we can also analyze how this relation changes when the
environment is driven (far) out of equilibrium and quantify the
violations of detailed-balance and the reciprocal relations in terms
of both excess dynamical activity ~\cite{lec05,mer05,gar09,ful13} and
probability currents. The latter result in the lack of an
action-reaction principle for the induced probe interactions
\cite{dzu03,hay06,bue08,ivl15}.  Section~\ref{sec:ex} exemplifies the
theoretical scheme with the help of two paradigmatic examples that can
explicitly be worked out.  First, we treat analytically a single probe
linearly coupled to a fluid of self-propelled particles. This toy
model clearly displays the breakdown of detailed balance and allows us
to touch on the scope of the notion of effective
temperature. Secondly, we employ Brownian dynamics simulations to
analyze the effective friction forces induced between two probes
suspended in a driven fluid. The numerical evaluation of our general
analytic expressions for the time-dependent friction matrix nicely
reveals the expected violations of the action-reaction principle, as
well as the possible appearance of negative mobility.  Finally, in
Sec.~\ref{sec:end}, we conclude with a summary and an outlook.

\section{General theory}\label{sec:theory} 
We consider a $d$-dimensional system made up of $M$ probe particles, with mass $m_i$ and positions $Q_i$, which interact with an environment, composed of $N \gg 1$ degrees of freedom denoted $x_k$. The environment is in contact with an equilibrium bath at inverse temperature $\beta$. The probes obey Newton's equation of motion
\begin{align}\label{probe}
&m_i \ddot Q_i = K_i(Q_i)+g_i(\{x_k\}, Q_i),  
\end{align}
where $g_i\equiv -\lambda\partial_{Q_{i}} U_i\left(\{x_k\},Q_{i}\right) $ is the interaction force between the probe $i$ and the environment, with $\lambda$ a small dimensionless parameter. All the other forces are incorporated in~$K$, which are (optional) direct interactions between the probes and additional external ones. Their specific form is irrelevant in the following. They are only required to be sufficiently confining so as to allow for a unique stationary state. Throughout the text, we use the shorthand $\{\dots_k\}$ to denote the entire set of degrees of freedom labelled by $k$. We assume that the environment evolves according to a Markovian stochastic dynamics, enjoying local detail balance. Hence, with respect to standard approaches, we lift such condition from the dynamics of the system to that of the environment. For concreteness, we can think of the overdamped Langevin equations
\begin{align}\label{env}
 \dot x_k=&\,\mu F-\lambda \mu \sum_{i=1}^M \partial_{x_{k}}  U_i+\sqrt{2\mu/\beta}\xi_k .
\end{align}
Here $F(\{x_k\})$ consists of inter-particle potential forces $-\partial_{x_k} V(\{x_k\})$, and external ones that may contain a non-potential driving $f(\{x_k\})$ setting the environment out of equilibrium. The $\xi$'s are centered Gaussian noises, white and uncorrelated.

Let $\{Y_i\}$ be the set of average positions around which the probes
fluctuate as a consequence of the interactions with the
environment. Here we are concerned with the fluctuations induced by
the presence of the environment, for which we seek a reduced
description. Namely, we aim at integrating out of \eqref{probe} the
environment coordinates by averaging the probe-environment coupling
with the appropriate distribution for $x_k$. We expect noise and
friction to emerge in this process, together with indirect forces
between the probes, mediated by the environment.  To this end, we
rewrite Eq.~\eqref{probe} as
 \begin{align}\label{ave}
m_i \ddot Q_i &= K_i+ \mean{g_i} + \eta_i, 
\end{align}
where we split the environment-probe coupling into a systematic part $\mean{g_i}$ and a random contribution $\eta_i \equiv  g_i -  \mean{g_i}$.
The former is defined as the mean force exerted by the environment on probe $i$, and reads
\begin{align}\label{aveg}
\mean{g_i(\{x_k\}, Q_i)} &\!\equiv\!  \int  \diff\{ x_k \} g_i(\{x_k\}, Q_i) \rho(\{x_k\}|\{Q_i\}),
\end{align}
with $\rho(\{x_k\}|\{Q_i\})$ the probability density of the environment conditioned by the probes being in positions~$\{Q_i\}$. We work under the usual assumptions made in the derivation of Langevin equations, i.e. a small variation of the probe momentum after a single particle-probe interaction (large mass difference), and a weak coupling between probes and environment. %[\textbf{needed for Gaussianity maybe:  a large number of interactions on the characteristic time scale on which the probe evolves ($N \gg 1$)}].
Under these conditions the fluctuations of probe $i$ around the preferred state $Y_i$ are small, and its force on the whole environment can be expanded to linear order in the displacement from $Y_i$:
\begin{align}
\lambda \sum_{k=1}^N \partial_{x_k} U_i& = \lambda \sum_{k=1}^N \partial_{x_k} U_i\bigg|_{Q_i=Y_i} \nonumber \\&
\quad -(Q_i(t)-Y_i) \sum_{k=1}^N \partial_{x_k} g_i\bigg|_{Q_i=Y_i}. \label{perturb} 
\end{align}
Here it is useful to regard $g_i$ as an external potential perturbing the environment, modulated in time via the protocol $Q_i(t)-Y_i$.  In view of \eqref{perturb}, it is then natural to express the conditional average \eqref{aveg} in in terms of unperturbed averages $\mean{\dots}^0$, corresponding to all probes sitting in the mean positions $\{Y_i\}$. To do so we make use of the response theory for perturbations about non-equilibrium states. The linear response formula in general reads \cite{bai13}
\begin{align}\label{resp}
\mean{\caA(t)} =& \,\mean{\caA(t)}^0 + \frac \beta 2 \sum_{j} \int_{t_0} ^t \diff s \,h_j(s) \nonumber\\
&  \times \left( \frac{\diff }{\diff s} \mean{\caB_j(s) ;\caA(t)}^0-\mean{L \caB_j(s); \caA(t)}^0 \right),
\end{align}
where $\caA$ is the observable of interest, $\caB_j$ are the perturbation potentials switched on at time $t_0$ and modulated in time through the protocol $h_j(s)$. The operator $L$ and the average $\mean{\dots;\dots}^{0}$ stand for, respectively, the backward generator of  the unperturbed dynamics and the connected average with respect to it. In \eqref{resp} the first integrand is the usual correlation of the observable with the entropy production, as appearing in the Kubo formula. The second one is a frenetic contribution that contains the excess dynamical activity, $L \caB_j$, caused by the perturbation. In equilibrium, they make equal contributions \cite{bai09}:
\begin{align}\label{equi}
 \frac{\diff }{\diff s} \mean{\caB_j(s) ;\caA_i(t)}^{\rm eq}=-\mean{L \caB_j(s); \caA_i(t)}^{\rm eq}.
\end{align}

Here we are interested in the response of $g_i(\{x_k\},Q_i)$ to the perturbations in \eqref{perturb}. Hence, with the identifications  ${\caA=g_i}$, ${\caB_j=   g_j}$  and ${h_j= Q_j-Y_j}$, \eqref{resp} becomes
%\begin{align}
%\langle g_i(\{x_k(t)\},& Q_i(t))\rangle- \mean{g_i(\{x_k(t)\}, Y_i)}^0 =\nonumber \\& \sum_{j=1}^M \frac \beta 2 \int_{t_0} ^t \diff s (Q_j(s)-Y_j) \times \nonumber\\
%& \bigg[ \frac{\diff }{\diff s} \mean{g_j(\{x_k(s)\}, Y_j); g_i(\{x_k(t)\}, Y_i) }^0 \nonumber \\ &-\mean{L g_j(\{x_k(s)\}, Y_j);g_i(\{x_k(t)\}, Y_i)}^0 \bigg],  \label{g2}
%\end{align}
\begin{align}
\mean{g_i(t)} &= \mean{g_i}^0  + \sum_{j=1}^M \frac \beta 2 \int_{t_0} ^t \diff s (Q_j(s)-Y_j) \nonumber\\
&\quad \times \bigg[ \frac{\diff }{\diff s} \mean{g_j(s); g_i(t) }^0-\mean{L g_j(s);g_i(t)}^0 \bigg],  \label{g2}
\end{align}
where $L$, the backward generator of the unperturbed dynamics of the environment, reads for~\eqref{env}:
\begin{align}
L=& \mu \underset{k=1}{\overset{N}{\sum}} \left[  F_k\partial_{x_k}- \lambda \sum_{i=1}^{M}\partial_{x_k} U_i \Big|_{Q_i=Y_i}\partial_{ x_k}+\frac{1}{\beta}\partial^{2}_{x_k}\right ] \label{L}.
\end{align}
The summands in~\eqref{g2} are the forces due to the linearized fluctuations of the probes around their preferred states.  Assuming that the environment was put in contact with the probes at time $t_0 = -\infty$, so that no correlation with the initial conditions is retained, an integration by parts yields
\begin{align}
\mean{g_i(t)}=&\,\mean{g_i}^0+ \sum_{j=1}^M\bigg[ G_{ij}(t)
% \nonumber\\&
-  \int_{- \infty} ^t  \diff s\, \zeta_{ij}(t-s)\dot Q_i(s) \bigg].\label{systematic2}
\end{align}
Here we defined the memory kernel
\begin{align}\label{zetaij}
\zeta_{ij}(t-s) &\equiv \frac \beta 2  \left(\mean{g_j(s); g_i(t) }^0 -\int_{-\infty}^s \diff u \mean{L g_j(u);g_i(t)}^0 \right), \end{align}
which enters both the friction and the statistical forces mediated by the environment,
\begin{align}\label{Gij}
G_{ij}(t)\equiv (Q_j(t)-Y_j) \zeta_{ij}(0),
\end{align}
including the ``self-interaction" (${i=j}$) and the forces between different probes  ($i\neq j$). 
 %We adopt a short hand notation hereafter for the correlations, as they depends only on $\{Y_i\}$, and on time just via the probability density of the environment.
Equation \eqref{Gij} establishes the connection between the friction kernel and the fluctuating statistical force, namely,
\begin{align}\label{sekimoto1}
&\partial_{Q_j} G_{ij}= \zeta_{ij}(0).
\end{align}
For $i\neq j$, Eq.~\eqref{sekimoto1} relates environment-mediated interactions to cross-friction between probes. It was proposed by De Bacco \emph{et al.} \cite{deb14} for equilibrium systems arguing on the basis of Onsager's regression principle. Here we gave a formal proof of this relation that extends its validity to nonequilibrium states. 

In equilibrium, where averages are denoted $\mean{\dots}^{\rm eq}$, the frenetic contribution can be eliminated in favor of the entropic term, according to~\eqref{equi}:
\begin{align}\label{equi2}
 \mean{g_j(s); g_i(t) }^{\rm eq} = -\int_{-\infty}^s \diff u \mean{L g_j(u);g_i(t)}^{\rm eq}.
\end{align}
We thus retrieve that the friction kernel is a symmetric matrix
\begin{align}
\zeta_{ij}(t-s)&= \beta \mean{g_j(s); g_i(t)}^{\rm eq} %\nonumber
%\\ &= \beta \mean{g_i(s); g_j(t)}^{\rm eq} 
=\zeta_{ji}(t-s), \label{sym}
\end{align}
since correlations are functions of $|t-s|$ only, thanks to time-reversal invariance. The symmetry \eqref{sym} translates into the condition $\partial G_{ij}/\partial Q_j=\partial G_{ji}/\partial Q_i$, which suffices to make $g_i$ derive from an effective (thermodynamic) potential $\caF(\{Q_i\})$. That such potential is the Helmholtz free energy of the environment,
\begin{align}
\caF\equiv - \frac 1 \beta \ln  \int  \diff \{x_k \} e^{-\beta( \lambda  \sum_{i=1}^N U_i+V)} ,
\end{align}
 is easily seen by introducing the Boltzmann factor in \eqref{aveg}:
 \begin{align}
\mean{g_i}^{\rm eq} &= - \int  \diff \{x_k \} \lambda \partial_{Q_i} U_i \, e^{-\beta(\sum_{i=1}^N U_i+V-\caF)}\nonumber \\
&= \frac 1 \beta  e^{\beta \caF}  \partial_{Q_i}  \int  \diff \{x_k \}  e^{-\beta( \lambda  \sum_{i=1}^N U_i+V)}\nonumber \\
&= \frac 1 \beta  e^{\beta \caF}  \partial_{Q_i}  e^{-\beta \caF}
= - \partial_{Q_i} \caF .
\end{align}
 This ensues the action-reaction principle for the fluctuating forces among probes. Contrarily, when the environment is driven away from equilibrium, \eqref{equi2} is not applicable in general, as frenetic and entropic terms remain distinct. Hence the reciprocal relations are not satisfied, $\zeta_{ij} \neq \zeta_{ji}$, which implies that the action-reaction symmetry is broken. 
 % [\textbf{$e^S$ breaks time-reversal invariance in the correlations, write that? Connection with last paper by Maes on statistical forces?}]

Now we turn to the random part of the interaction,
\begin{align}\label{noise}
{\eta_i \equiv  g_i(\{x_k\}, Q_i) -  \mean{g_i(\{x_k\}, Q_i)}}.
\end{align}
 It has zero mean by definition, and its two-times correlation is obtained again by application of the response formula \eqref{resp}, with $\caA= g_i g_j$,
\begin{align}
\mean{\eta_i(t) \eta_j(s)}&= \mean{g_i(\{x_k(t)\}, Q_i(t));g_j(\{x_k(s)\}, Q_j(s))} \nonumber \\
&\simeq \mean{g_i(t);g_j(s)}^{0} %+O(\lambda^3)
. \label{gg0}
\end{align}
The weak-coupling approximation allowed us to drop higher orders in $\lambda$, so that \eqref{gg0} simplifies to
\begin{align}\label{NEfdt}
\mean{\eta_i(t) \eta_j(s)}= \frac{2}{\beta} \zeta_{ij}(t-s) + \int_{-\infty}^s \diff u \mean{L g_j(u);g_i(t)}^0.
\end{align}
In general, the noise correlation depends explicitly on the excess dynamical activity of the environment,~$Lg_i$. Yet, in equilibrium, exploiting again the equality of the frenetic and entropic term,~\eqref{NEfdt} reduces to the FDT,
\begin{align}\label{fdt}
\mean{\eta_i(t) \eta_j(s)}^{\rm eq}= \frac{1}{\beta} \zeta_{ij}(t-s) .
\end{align}
Out of equilibrium~\eqref{NEfdt} cannot be simplified further in general, and the FDT~\eqref{fdt} is evidently broken, resulting in asymmetric noise cross-correlations.
%\begin{align}
%\mean{\eta_i(t) \eta_j(s)}&=  \mean{g_i(t);g_j(s)}^{0} \nonumber\\
%&\neq  \mean{g_j(t);g_i(s)}^{0}= \mean{\eta_j(t) \eta_i(s)}
%\end{align}
Such violation of the FDT appears more transparent if~\eqref{NEfdt} is written in terms of the state velocity of the environment, i.e.\ the vector
\begin{align}
v(\{x_k\}, \{Q_i\})\equiv  \frac{\jmath(\{x_k\}, \{Q_i\})}{\rho^0(\{x_k\}|\{Q_i\})},
\end{align}
with $\jmath$ the probability current of the environment, that vanishes identically in equilibrium. Even though it could be experimentally estimated \cite{bli07a, bat16, meh12, lan12, gom09},  it has been analytically solved only in few simple situations where the stationary distribution is known \cite{che08, ver11, bai16}. From the identity $L=L^*+2 v \cdot \nabla$ \cite{spe06, che08, bai13}, where $\nabla$ is the vector of partial derivatives $\partial_ {x_k}$, and $L^*$ is the adjoint of $L$---the forward generator of the dynamics of the environment---one can easily prove:
\begin{align}
\mean{L g_i(u);g_j(t)}^0=&% \mean{ \left( L^*+2 v\cdot \nabla \right)g_i(u);g_j(t)}^0 \nonumber \\
%&= \mean{ g_i(u); L g_j(t)}^0+2\mean{v \cdot \nabla g_i(u);  g_j(t)}^0\label{adjoint} \\
%&= \frac{\diff}{\diff t} \mean{ g_i(u); g_j(t)}^0 +2\mean{v \cdot \nabla g_i(u);  g_j(t)}^0\label{backward} \\
-\frac{\diff}{\diff u} \mean{ g_j(u); g_i(t)}^0 \nonumber \\&+2\mean{v\cdot\nabla g_j(u);  g_i(t)}^0 \label{ss}.
\end{align}
%where we used the definition of the adjoint and backward generator in Eq.~\eqref{adjoint} and Eq.~\eqref{backward}, respectively, and the time-translation invariance of steady-state correlations in Eq.~\eqref{ss}. 
Using Eqs.~\eqref{zetaij},~\eqref{NEfdt}~and~\eqref{ss} the broken FDT reads 
\begin{align}\label{vel}
&\mean{\eta_i(t) \eta_j(s)}= \frac{1}{\beta} \zeta_{ij}(t-s) + \int_{-\infty}^s \diff u \mean{v \cdot \nabla g_j(u);  g_i(t)}^0,
\end{align}
where the deviation from the equilibrium Kubo formula appears explicitly. 

In general, the noise $\eqref{noise}$ will not be Gaussian and thus the two-times correlation is not enough to fully characterize its statistics. Higher moments can be calculated with the same procedure, though, by successive application of the response formula \eqref{resp} together with the weak-coupling assumption.

Finally, we note that the restriction of time-independent mean states $\{Y_i\}$ can be easily lifted. If, instead, mean time-dependent trajectories $\{Y_i(t)\}$ are taken, our approach still holds with the caveat that the perturbation potentials, $ g_i(\{x_k\}, Y_i(s))$, now carry an explicit time dependence via $Y_i(t)$ (cf. Eq.~\eqref{perturb}). An extension of the response formula \eqref{resp} needs to be applied \cite{mae14}, which features $\{Y_i(t)\}$ as a quasi-static protocol, but the remaining procedure is very analogous. %[\textbf{write extra term?}]
Therefore, the theory naturally extends to probes that are, e.g., acted upon by external time-dependent forces, or in direct contact with the equilibrium bath, as well as with the environment. 
\section{Examples}\label{sec:ex}
In this section we present two explicative examples. First, we consider a single probe coupled linearly to a fluid of noninteracting self-propelled particles. Equation \eqref{zetaij} and \eqref{NEfdt} are calculated analytically and used to show the breakdown of \eqref{fdt}. Second, we show how to extract from Brownian simulations the friction memory kernel of two confined probes immersed in a stirred fluid. We prove numerically the breakdown of the reciprocal relations, that is the violation of the action-reaction principle for the fluid-mediated forces between the probes.
\subsection{One probe in an active fluid}\label{sec:active}
We consider a two-dimensional system ($d=2$) where a single probe under harmonic confinement,
\begin{align}\label{probelin}
K(t)=&-\kappa_Q \left(Q(t)-Y\right),
\end{align}
interacts via a harmonic potential $U$ (strength constant $\kappa$)
with an environment of active Brownian particles \cite{sol15}. The
latter are not mutually interacting but (internally) driven so that
they display a drift velocity of constant magnitude $v_{0}$ pointing
along the random particle orientation $n_k (t)$, i.e.
\begin{align}\label{fluid}
\mu F(\{x_k\})=v_{0}&n_k (t) \;.
\end{align}
Due to rotational Brownian motion, the unit vector $n_k (t)$ diffuses
with a persistence time $D_r$ \cite{ten09}:
\begin{align} \label{nn}
\left \langle n_k (t)n_{k'} (s)\right\rangle =\delta_{k k'}\, e^{-D_{r}\left|t-s\right|}.
\end{align} 
Therefore, \eqref{env} takes the simple form of an Ornstein-Uhlenbeck
process with an additional stochastic drift \cite{Gardiner}.
%The fluid equation is an Ornstein-Uhlenbeck (OU) process, but it admits a steady state only for the time-independent parameter $Y$. In order to write the coupling to the probe in function of the constant minimum $Y$, an exact expression can obtained summing and subtracting $\kappa\mu Y$ in \eqref{fluid}. Therefore \eqref{fluid} can be read as an O-U process plus a constant (in the fluid variables) perturbation. 
Thanks to the linearity of the system, the systematic part of the interaction,
\begin{align}\label{probe1}
\mean{g}=-\lambda\kappa \overset{N}{\underset{k=1}{\sum}}\left(Q-\left\langle x_k \right\rangle\right),
\end{align}
as well as the stochastic one,
\begin{align}\label{noise2}
\eta=-\lambda\kappa\overset{N}{\underset{k=1}{\sum}}\left( \left\langle x_k \right\rangle -x_k \right),
\end{align}
can be expressed analytically in terms of $Q$ and $Y$, only. Indeed, the terms in square brackets in \eqref{g2}, corresponding to the response function to a constant force, 
\begin{align}\label{response}
&\frac{N \beta \lambda^{2}\kappa^{2}}{2}\bigg[\frac{\diff}{\diff s}\left\langle x_k (s);x_k (t)\right\rangle ^{0}-\nonumber \\ & \qquad v_{0}\left\langle n_k (s);x_k (t)\right\rangle ^{0}+ \lambda\kappa\mu\left\langle x_k (s);x_k (t)\right\rangle ^{0}\bigg],
\end{align}
contain simple correlation functions of the unperturbed Ornstein-Uhlenbeck steady state. From \eqref{zetaij} we thus obtain the friction kernel 
\begin{align}\label{frictionlin}
\zeta\left(t-s\right)=N\lambda\kappa e^{-\lambda\kappa\mu\left(t-s\right)},
\end{align}
showing that dissipation happens on the characteristic timescale it
takes the active particles to relax in the coupling potential~$U$.  In
contrast, the energy input due to the noise \eqref{noise2} is found to
occur on multiple timescales,
\begin{align}\label{FDTlin}
\left\langle \eta\left(s\right)\eta\left(t\right)\right\rangle=&\frac{1}{\beta}\zeta(t-s)+\frac{1}{2}\frac{N\lambda\kappa v_{0}^{2}}{\left(\lambda\kappa\mu\right)^{2}-D_{r}^{2}} \nonumber \\ 
& \times \left[\lambda	\kappa e^{-D_{r}\left(t-s\right)}-\frac{D_{r}}{\mu}e^{-\lambda \kappa\mu\left(t-s\right)}\right].
\end{align}
The disparity of the time scales for noise and friction entails the
breakdown of the FDT, as predicted by \eqref{NEfdt}. One may try to
mend it by introducing an effective temperature \cite{cug11} via
\begin{equation}
\beta T_{\rm eff}(\tau) = 1 +\frac{\beta v_{0}^{2}}{2\mu D_{r}} 
 \left(1-\frac{\lambda\kappa \mu}{D_{r}} e^{-D_{r}\tau}\right) + {\cal
 O}(\lambda^2)\;. \label{Teff}
\end{equation}
Thereby, the FDT \eqref{fdt} is formally restored, albeit with the
time-dependent function $T_{\rm eff}(\tau)$ replacing the constant
bath temperature $1/\beta$.

The deviation from equilibrium is seen to
be governed by the two dimensionless numbers $\beta v_{0}^{2}/2\mu
D_{r}$ and $\lambda\kappa\mu/D_r$. For $\lambda\kappa\mu/D_r\to0$,
the temperature renormalization becomes time-independent and
independent of the weak-coupling parameter $\lambda$---it thus acquires the status of thermodynamic temperature.  One can then
justly say that the probe acts as an ideal measurement device for the
constant effective temperature 
\begin{equation}
T_{\rm eff} \sim \beta ^{-1} +\frac{v_{0}^{2}}{2\mu D_{r}} 
  \label{Teff2}
\end{equation}
of the active fluid itself, which coincides with the known value for a
suspension of free active particles \cite{yan14, tak15}.  The strength
of the temperature renormalization is controlled by the Pecl{\'e}t
number $v_0(\mu D_r/\beta)^{-1/2}$ that weighs the relative
importance of ballistic versus (translational and rotational)
diffusive motion \cite{bec16}.

To first order in $\lambda$, Eq.~\eqref{Teff} exhibits a crossover
from a short-time temperature to a long-time temperature. Moreover, $T_{\rm eff}$
can no longer be interpreted as a property of the particle bath alone,
but characterizes its interaction with the embedded probe.  In fact,
the ratio $\lambda\kappa\mu/D_r$ can be interpreted as a measure for
the interference of the coupling potential with the persistence of the
active particles motion.  We expect this particular feature to carry
over to more general (strongly interacting) systems, where it would
not be accessible within the weak-coupling formalism, however. The
physical picture is that the apparent thermalization at the constant
effective temperature \eqref{Teff2} takes some finite time to
happen. In our toy model, this ``equilibration time'' is given by the
rotational diffusion time of the active particles; i.e., the active
motion of the bath particles can only be subsumed into an enhanced
fluid temperature once it has lost its orientational persistence. This
very plausible condition has been pointed out before (e.g. in \cite{sol15c}), albeit
not for the time domain. If \eqref{FDTlin} is extrapolated to
values of the dimensionless coupling strength on the order of one, the
temporal growth of the corresponding effective temperature takes the
form
\begin{equation}\label{Teffcrossover}
\beta T_{\rm eff}(\tau)  \stackrel{\lambda\kappa\mu \approx
  D_r}{\approx} 1 + \frac{\beta
  v_{0}^{2}}{4\mu D_{r} }\left(1+\tau D_{r} \right)  \;.
\end{equation}
It may tentatively be interpreted as an indication of the onset of
strong interactions and collective effects, such as a clustering of
the bath particles around the probe, which would entail a progressive
heating of the probe. While quantitatively inaccessible to the
weak-coupling formalism, corresponding observations have indeed been
made in numerical simulations \cite{yan14, fod16}.

Summing up, we arrived at the generalized Langevin equation for the
probe,
\begin{align}\label{probfinlin}
M\ddot{Q}(t)=&-\kappa \left(Q(t)-Y\right)\nonumber \\&-\int_{-\infty}^{t}\diff s\, \zeta(t-s)\dot{Q}(s)+\eta\left(t\right),
 \end{align}
 where the friction memory kernel and the noise covariance are given
 by \eqref{frictionlin} and \eqref{FDTlin}, respectively. Note that, since $n_k$ is not Gaussian in general,
 \eqref{noise2} is not Gaussian either. Nevertheless, in view of the
 central limit theorem, the probability distribution of $\eta$
 converges to a normal one for $N \gg 1$, $\{x_k\}$ being independent
 identically distributed random variables.

\subsection{Two probes in a stirred fluid}
\begin{figure}[t]
\begin{center}
\includegraphics[width=0.45\textwidth]{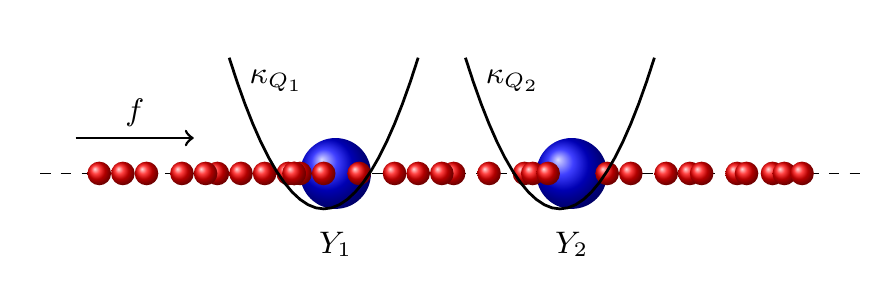}
\end{center}
\caption{Schematic illustration of the simulated system, composed of
  $N+2$ soft spheres in one spatial dimension with periodic boundary
  conditions. The probes (blue) have average positions $Y_i \sim
  Y_i^*+ \jmath \int_0^\tau \diff \tau \zeta_{ii}(\tau)/\kappa_{Q_i}$
  resulting from the balance of the drag force due to the steady
  current $\jmath$ of fluid particles (red) and to the harmonic
  confinement with minimum in $Y_i^*$ and stiffness $\kappa_{Q_i}$
  ($i=1, 2$). }
\label{fig:drawing}
\end{figure}
We consider a one-dimensional system ($d=1$) consisting of $M=2$ probes under
harmonic confinement and $N=100$ fluid particles moving freely in a
periodic domain $x_k \in [0,L]$, as sketched in
Fig. ~\eqref{fig:drawing}. The fluid is driven out of equilibrium by
an external constant force $f$ that induces a net particle current
$\jmath$ thanks to the periodic boundary conditions. The fluid
particles interact (mutually and with the probes) through the same
soft repulsive potential $V\equiv U$, such that they experience the
total force
\begin{align}
&F_k= f- \sum_{k'=1}^N\frac{\partial V_{k'} (x_k,x_{k'})}  {\partial x_k}.
\end{align}
 In the following we present results obtained by using the Gaussian potential
 \begin{align}
 V_{k'}(x_k,x_k')= e^{-\frac{1}{2 \sigma^2}{(x_k-x_{k'})^2}} \;,
 \end{align}
but we have checked numerically that anharmonic potentials lead to
qualitatively similar results. In particular, we have calculated the
time-dependent entries of the friction kernel $\zeta_{ij}$ from
formula \eqref{zetaij} for various values of the external driving
$f$. This was done by letting the fluid relax from an initial uniform
density, fixing the probes in their preferred positions $Y_i$, and
then performing the steady-state averages in \eqref{zetaij} over $2
\times 10^4$ independent simulation runs of duration $\caT=10^3$.

\begin{figure}[t]
\begin{center}
%\begin{tikzpicture}[scale=1]
% \pgfplotsset{legend pos=north east}
%\begin{axis}[
%xlabel={$\tau/\caT$},ylabel={$\zeta_{ii}(\tau)$}, ylabel near ticks]
%\addplot [black,very thick] table [x={t}, y={bho1}] {Frictionii.dat};
% \addlegendentry{$\zeta_{11}$, $f=0$}
%\addplot [dashed, red,very thick] table [x={t}, y={bho2}]{Frictionii.dat};
% \addlegendentry{$\zeta_{22}$, $f=0$}
%\addplot [dashed,blue,very thick] table [x={t}, y={bho3}] {Frictionii.dat};
% \addlegendentry{$\zeta_{11}$, $f=1$}
%\addplot [solid,blue,very thick] table [x={t}, y={bho4}] {Frictionii.dat};
% \addlegendentry{$\zeta_{22}$, $f=1$}
%\addplot [dotted,blue,very thick] table [x={t}, y={bho5}] {Frictionii.dat};
% \addlegendentry{$\zeta_{11}$, $f=2$}
%\addplot [dashed,yellow,very thick] table [x={t}, y={bho6}] {Frictionii.dat};
% \addlegendentry{$\zeta_{22}$, $f=2$}
%\addplot [dashed,green,very thick] table [x={t}, y={bho7}] {Frictionii.dat};
% \addlegendentry{$\zeta_{11}$, $f=3$}
%\addplot [dashed,purple,very thick] table [x={t}, y={bho8}] {Frictionii.dat};
% \addlegendentry{$\zeta_{22}$, $f=3$}
%\end{axis}
%\draw[thick,->](2.5,1.8)--(1.5, 0.8);
%\draw(2.3,1.6) node[above] {$f$}; 
%\end{tikzpicture}
\includegraphics[width=0.45\textwidth]{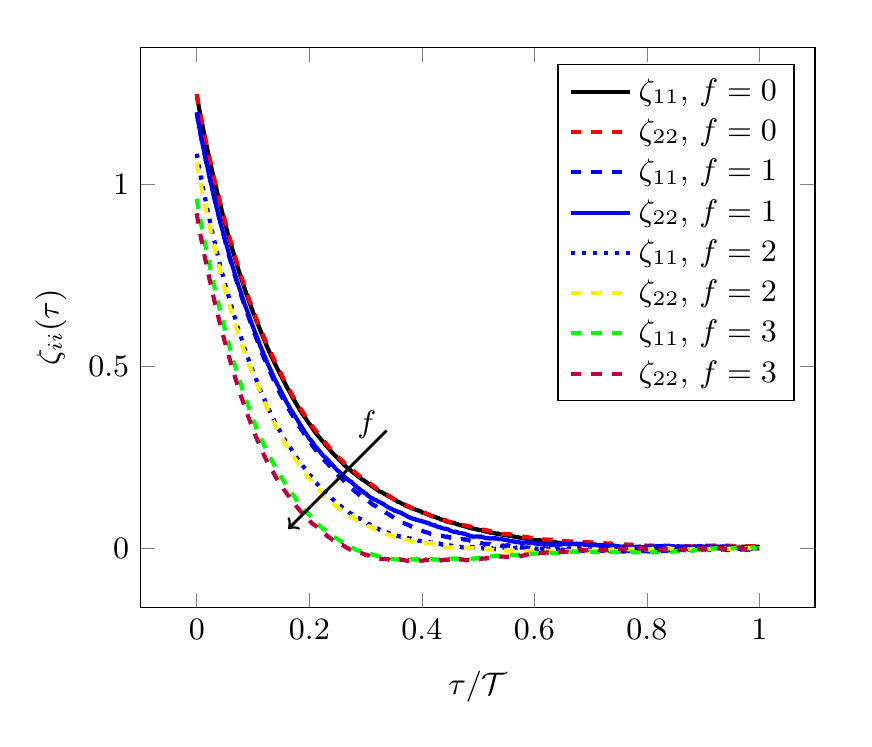}
\caption{Diagonal elements of the friction kernel $\zeta_{ij}(\tau)$ as function of time $\tau=t-s$, obtained by numerical evaluation of \eqref{zetaij} in Brownian dynamics simulations, for various values of the nonequilibrium driving force $f$ and $\beta=1$, $\mu=1$,  $\sigma=1$. }
\label{fig:zetaii}
\end{center}
\end{figure} 

\begin{figure}[t]
\begin{center}
%\begin{tikzpicture}[scale=1]
% \pgfplotsset{legend pos=south east}
%\begin{axis}[
%xlabel={$\tau/\caT$},ylabel={$\zeta_{ij}(\tau)$},ylabel near ticks]
%\addplot [black,very thick] table [x={t}, y={boh1}] {Friction.dat};
% \addlegendentry{$\zeta_{21}$, $f=0$}
%\addplot [dashed, red,very thick] table [x={t}, y={boh2}]{Friction.dat};
% \addlegendentry{$\zeta_{12}$, $f=0$}
%\addplot [dashed,blue,very thick] table [x={t}, y={boh3}] {Friction.dat};
% \addlegendentry{$\zeta_{21}$, $f=1$}
%\addplot [solid,blue,very thick] table [x={t}, y={boh4}] {Friction.dat};
% \addlegendentry{$\zeta_{12}$, $f=1$}
%\addplot [dotted,blue,very thick] table [x={t}, y={boh5}] {Friction.dat};
% \addlegendentry{$\zeta_{21}$, $f=2$}
%\addplot [dashed,yellow,very thick] table [x={t}, y={boh6}] {Friction.dat};
% \addlegendentry{$\zeta_{12}$, $f=2$}
%\addplot [dashed,green,very thick] table [x={t}, y={boh7}] {Friction.dat};
% \addlegendentry{$\zeta_{21}$, $f=3$}
%\addplot [dashed,purple,very thick] table [x={t}, y={boh8}] {Friction.dat};
% \addlegendentry{$\zeta_{12}$, $f=3$}
%\end{axis}
%\draw[thick,->](1.8,4)--(1.,5.3);
%\draw(1.2,4.2) node[above] {$f$}; 
%\draw[thick,->](0.8,2)--(1.3,1.75);
%\draw(1.2,1.2) node[above] {$f$}; 
%\end{tikzpicture}
\includegraphics[width=0.45\textwidth]{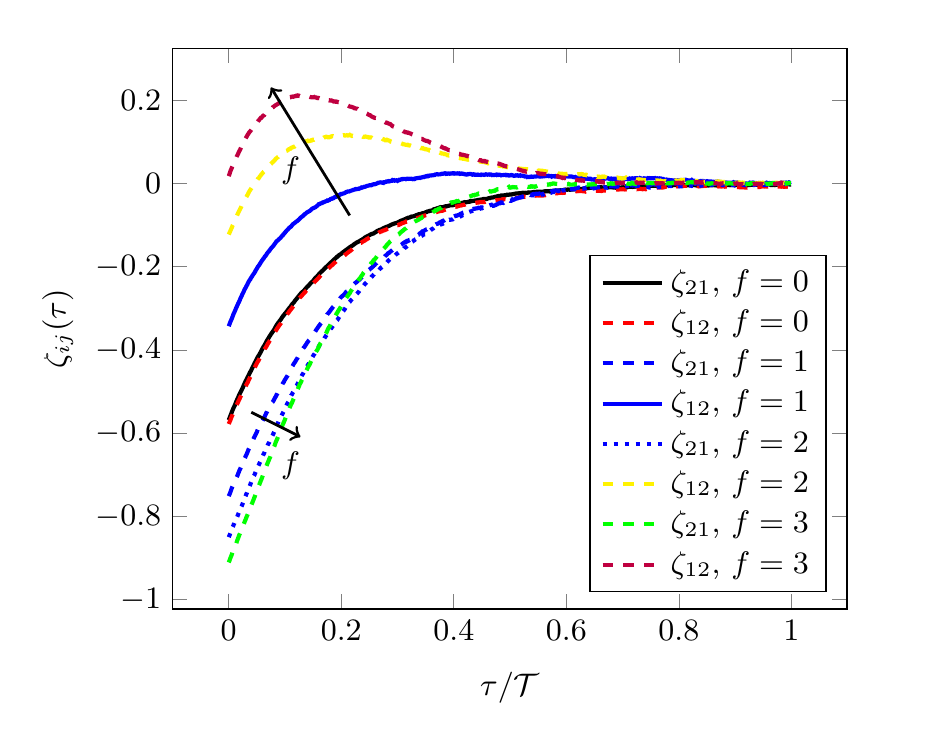}
\caption{Off-diagonal elements of the friction kernel
  $\zeta_{ij}(\tau)$ as function of time $\tau=t-s$, obtained by
  numerical evaluation of \eqref{zetaij} in Brownian dynamics
  simulations for various values of the nonequilibrium driving force
  $f$ and $\beta=1$, $\mu=1$, $\sigma=1$.}
\label{fig:zetaij}
\end{center}
\end{figure} 

For $f\to0$, equilibrium conditions are recovered.  The diagonal
elements $\zeta_{ii}$ of the friction kernel are positive
and exhibit a monotonic time dependence. The two off-diagonal elements
$\zeta_{12}$ and $\zeta_{21}$, which quantify the mutual frictional
forces between the probes, coincide. As expected, they are negative
and decay to zero at late times. Their negative sign can be understood
on the basis of global momentum conservation. For example, consider
the drag force that probe $1$ exerts on probe $2$,
%
%\begin{figure}
%\includegraphics[scale=0.50]{f=0.png}
%\end{figure}
\begin{align}
F_{1\rightarrow2}^{\rm drag}=-\int_{-\infty}^t \diff s \zeta_{21}\left(t-s\right)\dot{Q}_{1}\left(s\right) \nonumber .
%\\ F_{2\rightarrow1}^{drag}=-\int\zeta_{12}\left(t-s\right)\dot{Q}_{2}\left(s\right)
\end{align}
It is easy to convince oneself that, given the configuration sketched
in Fig.~\eqref{fig:drawing}, a positive velocity $\dot Q_1$ will on
average cause a positive displacement of the fluid particles
surrounding probe 1. Such perturbation spreads along the coordinate
axis reaching probe 2, ultimately resulting in a positive momentum
transfer $F_{1\rightarrow2}^{\rm drag} >0$. This suggests that
$\zeta_{21} \leqslant 0$ for all times.
%Suppose that $Q_{1}<Q_{2}$ (as well as in the simulations). If $Q_{1}$ is increasing its position $\dot{Q}_{1}\left(s\right)>0$, it will produce a perturbation that will propagate itself to $Q_{2}$ pushing it in the positive direction. It requires $F_{1\rightarrow2}^{drag}>0$ and it is possible only if $\zeta_{21}\left(t-s\right)<0$ (in purple in the plot). On the other hand, if $Q_{1}$ is decreasing its position $\dot{Q}_{1}\left(s\right)<0$, it will produces a perturbation that will propagate itself to $Q_{2}$ pulling it in the negative direction (indeed if the first probe moves away it creates a ``hole" in the fluid that will attract the second. At equilibrium this mechanism describes quite well why the friction is negative. 

In contrast, with increasing nonequilibrium force $f>0$, we observe a
qualitative modification of the diagonal and non-diagonal elements of
$\zeta_{ij}$, as exemplified in Figs.~\eqref{fig:zetaii} and
\eqref{fig:zetaij}, respectively.  The diagonal elements $\zeta_{ii}$
develop a non-monotonic time dependence and eventually turn
negative. Physically, this corresponds to a viscoelastic recoiling of
the individual probe particles. A more dramatic, genuinely
non-equilibrium effect is found for the off-diagonal elements
$\zeta_{i\neq j}$. As revealed by Fig.~\eqref{fig:zetaij}, the
presence of a nonequilibrium flux in the bath breaks the symmetry of
the friction matrix so that $\zeta_{i j}\neq \zeta_{ji}$, with
$|\zeta_{21}|$ ($|\zeta_{12}|$) larger (smaller) with respect to
equilibrium. Such an effect arises whenever a spatial asymmetry is
imposed on top of broken detailed balance. Our periodic system is
always spatially asymmetric unless $Y_1-Y_2 = L/2$. Specifically, in
the simulations, the probe reference positions are set to $Y_1\simeq
L/3 < Y_2\simeq L/2$, and, for convenience, the trap stiffnesses
$\kappa_{Q_i}$ are chosen large enough to make the position $Y_i$
almost coincide with the trap minimum $Y_i^*$. By increasing $L$, we
checked that interactions with the periodic image particles are
negligible.  We conclude that global momentum conservation does not
hold any more when the fluid dynamics becomes dissipative.  This can
be attributed to the asymmetric propagation (due to the current
$\jmath$) of fluid perturbations. Namely, downstream propagation is
progressively enhanced by increasing $f$, while upstream propagation
is suppressed. As a result, the influence of probe 1 (2) on probe 2
(1) gets stronger (weaker) as we increase the driving. As for the diagonal elements, the
sign of $\zeta_{12}(\tau)$ is transiently reversed. More remarkably, for
sufficiently large values of $f$, the response coefficient of probe 2
to a uniform motion of probe 1, namely $-\int_{0}^{\infty} \diff \tau
\zeta_{12}(\tau)$, turns negative. In contrast to the mentioned transient
elastic recoil embodied in the diagonal terms $\zeta_{ii}$, this kind
of ``absolute negative mobility'' \cite{eic02,ros05,mac07,gho14} is strictly
forbidden in equilibrium, where dissipative transport coefficients
depend only on the (positive) entropy production but not on the
dynamical activity \cite{bae13}.

\section{Conclusion}\label{sec:end}
Employing nonequilibrium linear response theory we have derived
generalized Langevin equations for probe particles interacting with a
driven environment.  The latter was described by an explicit
interacting many-body theory for overdamped colloidal particles
representing, e.g., active or sheared colloids. More generally, they
can be understood as a set of mesoscopic degrees of freedom. Also, the
theoretical framework developed above can be easily adapted to cope
with different sources of nonequilibrium (other than the
nonconservative force $f$), such as a nonuniform bath temperature
field $\beta(x_k)$.

When only conservative forces are present, our theory correctly
reproduces the expected equilibrium properties, i.e. it fulfils the
FDT and conforms to Onsager's regression principle relating the
fluctuations of statistical forces to the memory kernel. In general,
it extends the Langevin approach into the nonequilibrium realm,
predicting the violation of the FDT and the action-reaction law for
the fluctuating effective forces. The breaking of these dynamical
symmetries is traced back to the mismatch between the excess entropy
and dynamical activity induced by probes fluctuations around their
preferred states or, equivalently, to the existence of dissipative
currents in the environment. We have shown that these phenomena appear
already in simple systems, unless special symmetries are
present. Namely, noise and friction felt by a single probe in an
active medium do not obey the FDT, except if the
relaxation timescales of system and fluid are properly tuned ---in
which case a constant effective temperature can be defined. Also, the
cross-frictions between two confined probes in a stirred periodic
fluid are dissimilar, and even change sign with respect to
equilibrium, whenever the probe reference positions break the spatial
symmetry.

The theory allows to obtain quantitative information about the
 parameters of the environment from measuring average
properties of the probes. For example, from \eqref{frictionlin} and
\eqref{FDTlin} ---which are accessible by measuring, e.g., the
spectral density of the probe fluctuations in the trap and its
response to a small external kick--- the values of the relaxation
times $\mu \kappa$ and $D_r$ can be inferred. Vice versa, one may even
speculate that some mesoscopic parameters (e.g., $\zeta_{ij}(0)$)
might be fixed at will by properly designing the non-conservative
driving. It is in principle feasible since formal procedures are
available \cite{che15} which determine an appropriate environment
dynamics conditioned on prescribed mean values (e.g., those entering
\eqref{zetaij}).

Finally, a remark on the status of the approximation of weak coupling
to the nonequilibrium environment seems in place. In a particle-based
theory like the one we employed, this approximation is explicitly
enforced by introducing a small coupling constant $\lambda$. 
%Due to the global coupling to the fluid particles,
Physically, the appropriate values $\lambda$ may depend on the average number of bath particles the probes interact with. This should be
clear from the example in Sec.~\ref{sec:active}, where the limit $N
\to \infty$ produces an unphysical divergence of friction and noise
strength if $\lambda$ is not properly scaled.  However, in practical
applications, the weak coupling is often a dynamical, \emph{emergent}
property, resulting from the scale separation between the
probe-particle system and the environment.  For example, colloidal
particles suspended in simple fluids are well described by a linear
hydrodynamic theory, although the micro-dynamics of the fluid
molecules is highly nonlinear. This feature is expected to be robust
and to survive even far from equilibrium, as long as the driving
energy input does not exceed the bath thermal energy
\cite{fal16b}. Indeed, the peculiar feature of a time-dependent noise
temperature, discovered within the weak-coupling approach, above, was
already explicitly demonstrated (and its time-dependence analytically
computed) in this setting \cite{fal14}.

Recently, new theoretical investigations \cite{sei16,kru16} have been
spurred by a surge of experimental interest in systems with strongly
coupled components, such as in active nonlinear micro-rheology
\cite{gom15}, single-molecule (force spectroscopy) experiments
\cite{ott13}, work extraction from active fluids \cite{dil10}. Hence,
it would be desirable to extend the above analysis to different
dynamical descriptions of the environment, i.e. in terms
(hydrodynamic) fields or discrete-state variables. This will possibly
provide more versatile formal tools to account more reliably for the
weak coupling and to address the strong coupling problem in a larger
variety of stochastic systems.

\subsubsection*{Acknowledgment}
G. F. and K. K. acknowledge funding by the Deutsche Forschungsgemeinschaft (DFG). S.S. acknowledges funding by International Max Planck Research Schools ({IMPRS}).

\bibliography{IntBM_arxiv}
\bibliographystyle{apsrev4-1}

\end{document}